# TOWARDS A QUANTUM APPROACH TO CELL MEMBRANE ELECTRODYNAMICS

Jacques Breton, 9 Avenue de Gradignan, 33600 Pessac, France
Vincent Breton, CNRS-IN2P3, LPC Clermont, Université Blaise Pascal, CNRS/IN2P3, Aubière, France

Corresponding author : Jacques Breton, jacqueton@free.fr

**ABSTRACT** The ultimate active constituents of the living medium, membranes, ions and molecules, are **at the level of the nanometer**. Their interactions thus require a **quantum processing**. The **characteristic Action A**, linked to the « quantum objects » : ions, radicals, water molecule… of the living medium, has an average value of $A \approx 14.10^{-34}$ J.s or $A \approx 2h$. It is **thus strictly impossible** to formulate a realistic "classical" theory of membrane electrodynamics. T**he transfer of the sodium ion – among others – could be then ensured under the action of a Tunnel effect, (with Hartman's mechanism) specific to the channel and the ion transferred.**

The cell, structural unit common to all living organisms, is dependent for its existence and longevity upon its surrounding membrane. The ultra-plasticity of this membrane allows the cell to constitute the sub-units necessary to its operation (cytoskeleton, endoplasmic reticulum, Golgi apparatus, vacuoles….) and to manage all of the intercellular relationships, as well as the obtaining, assimilation and excretion operations of different metabolites. This involves delicate "mechanics" ensured and checked by various proteins (clathrins, dynamins, kinesins…. ), actual molecular motors operating at the micrometric scale or at hundreds of nanometers on the objects mentioned. In all cases, the plastic properties of a "quasi-liquid" membrane are always utilized (1) (2) (3) (4). Beyond the variety of its constituents and their "mechanical" properties, however, and at the most fundamental level, this living medium dominated by the properties of the water molecule, is in fact of an electro-ionic nature. Housing numerous structural discontinuities, it is necessarily the place of a complex electrodynamics, which is the key and ultimate driver of its harmonious functioning.

From these discontinuities result in particular gradients of concentration of a few ions or essential radicals such as $Na^+$, $K^+$, $Mg^{++}$, $Ca^{++}$, $H^+$, $OH^-$, $Cl^-$, $CO_3^{--}$, $PO_4^{--}$, $Fe^{++}$, $Fe^{+++}$, to which are added some more "exotic" but nonetheless important ions. It is imperative to take note that even though these ions are "free" in the locality of the membrane, they are in fact "confined" within this locality. This confinement is essentially the result of the Coulombian attraction of the radicals ($-CO_2^-$) carried by the Cell Coat glycoproteic chains, "lining" the membrane's exterior. Though we are of course dealing with an overall electrically neutral space, we are, however, in the presence of a highly polarized space, resulting from very significant localized charges. The immediate result is therefore the existence of local electrical fields, of great value, presenting a complex distribution and strong gradients. The laws of electrokinetics make these the "motors" of intense exchanges across the membranes, along them and inside of them; they verify and "manage" in particular the transmembranous exchanges via the ionic channels, and the movement, ensured by the molecular motors, of diverse organelles in the very interior of the cell; act therefore also upon the conformation and the reactivity of molecules or molecular clusters qualified for ordering or maintaining these exchanges or movements. They are also actors in the dispatching of nervous influx, motors of biochemical reactions between ions, molecules and structures. They are capable of maintaining the cohesion and autoreparation of the membrane, unstable by nature, by means of the very intense forces of electrostriction that the enormous radial electric field ($\geq 10^7$ V.m$^{-1}$) imposes on an intrinsically insulating structure

Finally, omnipresent electrical field gradients will act powerfully upon all polarizable molecules, and particularly upon the water molecule, which is neutral but equipped with a very strong dipolar moment, thus ensuring and modulating their movements, their transfer (via the aquaporins, among others) and their reactivity.

This being the case, one can no longer ignore that the mechanisms at work are situated at the level of the nanometer, often even beneath, and no longer at the level of our habitual "macroscopic" space. ,

A result is that at this scale, separated from our usual scale by 9 or10 orders of size, the entire operation of the cellular "machinery"(just as for the semiconductor mediums) depends certainly upon electrokinetic processes <u>fundamentally different</u> from the "macroscopic" processes. The membrane dynamics, which results from this operation, therefore rests upon these "non-classical" processes. Two local mechanisms, a depolarization of the membrane and the opening of the channels, on the other hand the permanent output of the $K^+$ ion of the cell at rest, and the "return" of the $Na^+$ and $K^+$ ions at the moment of the return of the membrane to the "resting" state following a depolarization-excitation.

However, contrary to the rules of classical electrokinetics, a number of these operations exert themselves <u>against</u> large potential gradients [ 5 ] [ 9 ]. The result is a major difficulty in continuing to propose "classical" mechanisms for transferring the concerned ions; the following citation unequivocally translates this situation: " Ions on one side of the membrane bind to sites within the protein and become temporarily occluded (trapped within the protein) before being released to the other side, but details of these occlusion and de-occlusion transitions <u>remain</u> <u>obscure</u> for all P-type Atp-ases" [6].

It therefore seems imperative to reconsider all the mechanisms at cause by means of a "non-classical" approach, the <u>quantum</u> approach being here the only operational and pertinent approach, with regard to the dimensions of the objects interacting [10] [11] [12]. This new hypothesis may be thus expressed as follow:

"the transfer of the sodium ion – among others – could take place across the corresponding ionic channel under <u>the action of a Tunnel effect specific to this channel and to the ion transferred</u>, with the minimum consumption of energy belonging to this mechanism, and, conforming to observation, <u>against the transmembrane potential gradient</u>. Such a process is perfectly compatible with the quantum characteristics of the sodium ion and the characteristics of the Tunnel effect, and <u>the Hartman's mechanism</u> [13]. We point out that the first observation of "Hartman effect" was done with photons [14] and later confirmed on particles treated in a relativistic way. The quantum object with non zero mass interacts as a gaussian wave packet with the quantum barrier which has low transmitivity and acts as a high frequency filter. The observed transfer time of the wave packet / quantum object becomes quicly independant of the barrier thickness, as if the transfer was taking place at a "supraluminic" speed. The "stationary phase" method applied to the incoming and outcoming wave packets provides the right theoretical framework for the description of this phenomenon which of does not violate laws of relativity as the energy associated to the object does not propagate quicker than the speed of light [15] [16]. Such mechanisms would grant to the process in question the speed and the perfect efficiency necessary for the development of the ionic distributions and for the preservation of their specific characteristics. The ATP-to-ADP conversion would thus free an energy essentially assigned to the other processes, in particular the functioning of the molecular motors and the return to a state of rest of the membrane. This same Tunnel effect would allow the potassium ion, which is of crucial importance for the cellular dynamic, to execute the entries-exits necessary to the exercise of this dynamic, while the water molecule, nonetheless essential, could circulate efficiently via the aquaporins. Other ionic or molecular transfers could, depending upon their characteristics, use the same path with the same specificity, the same efficiency and the same precious economy of energy ".

This is no longer an ad hoc hypothesis, but a reminder of the obligation of living organisms to respect a fundamental characteristic of the quantum level at which are located mechanisms which it implements. <u>Nothing indeed permits one to think that this medium could avoid, for an unspecified reason, the strict application of the quantum laws which concern it and on which it depends</u>. It thus seems imperative to reconsider all of the mechanisms at cause by means of a "non-classical" approach, the <u>quantum</u> approach being then here <u>only operational and relevant</u>, compared to <u>dimensions</u> of the objects in interaction [10] [11] [12].

The very basic reminder, which will follow, therefore constitutes this <u>first "non-classical" approach</u> to membrane dynamics.

REMINDERS OF DIMENSIONAL ANALYSIS

We have at our disposal a basic universal system of basic sizes which are the mass M, length L and time T. Furthermore, the frequency ν has a dimension of $|ν| = T^{-1}$ while energy E is defined by $|E| = M L^2 T^{-2}$.
From the law of the photoelectric effect $E = hν - W$ we obtain the dimension equation of h
$$|h| = M L^2 T^{-1}$$

But h, Planck constant, essentially concerns the atomic space, of which it characterizes <u>all</u> of the properties. Its value ≈ 6 . 10 $^{-34}$ J.s is perfectly representative of the <u>scale</u> of the objects in which it intervenes (atoms, ions, molecules…).

Faced with the obligatory choice between "classical" <u>or</u> "quantum" treatment, it is essential to define a generic term, having the <u>dimension</u> of **h**, for all physical size susceptible to intervening at <u>this atomic scale</u>.

But in classic mechanics the Lagrangian formalism defines under the name of "Action" a space integral of momentum and/or a temporal integral of energy which disposes precisely of the dimension h. One thus refers to as Action, abbreviated A, <u>all physical quantity</u> of the dimension $M L^2 T^{-1}$.

Let us also remember that the three fundamental physical sizes constantly used, which are mass, length and energy, are <u>independent sizes</u>. There is therefore <u>necessarily</u> a <u>single</u> monomial relationship between these three quantities, having <u>the dimension of an action</u>.

Let us therefore note here :  $(M L^2 T^{-1})^2 = (M L^2 T^{-2}) (M) (L^2)$

Or, finally  $\boxed{|\text{Action}|^2 = |\text{Energy}| . |\text{Masses}| . |\text{Length}|^2}$

Besides, observations and measurements prove that the Planck constant characterizes the scale of events arising from <u>sole quantum physics</u>. It can therefore constitute the <u>natural</u> "standard" of all action linked to a physical system.

For this reason and as a result of this fact, <u>it strictly defines</u> the domain of validity of "traditional theories". These can only remain pertinent, and therefore can only be used as a <u>first</u> <u>approximation</u>, if <u>all the sizes</u> in question, of the type "Action", are <u>very large</u> in front of h.

If it happens that the action linked to physical quantity is of a value comparable to h, experience confirms that it is then <u>prohibited</u> to disregard the quantum effects linked to this quantity, and to rest upon the "<u>classical</u>" approximation.

One can thus condense this rule

"Action about h <=> quantum  treatment

Let us remember here that a multitude of experiences, of which the precision keeps growing, have confirmed this rule, <u>never</u> put at fault; classical theories and the methods of application which result from them always turn out to be <u>completely unfit</u> for predicting and expressing the physical content or the properties of objects <u>at the scale of h</u>

The question thus asked here is: "which is the value of the characteristic Action of the mechanisms taking place within the cell membrane ?

The three quantities at cause are always a mass, a length and an energy; these are the intrinsic characteristics of the "quantum objects" which constitute the ions or molecules, <u>active factors</u> in the membrane electrodynamics.

- Here, the "mean" mass M  is therefore that of fundamental ions, sodium or potassium, or that of a water molecule, all objects subject here to complex forces from which their movements or interactions result.

In the case of the ion $Na^+$, this value is in the amount of  $23.10^{-3}$ kg / $6.10^{23}$ ≈ $4.10^{-26}$ kg

- The length L at cause (size <u>related</u> <u>to the quantum object</u> which one seeks the Action) is therefore <u>necessarily</u> the diameter of the ion or of the molecule considered ; it is in the amount of 0,1 nanometer, or $10^{-10}$ meter.

- The energy E to take into consideration results on average from only the thermal agitation of these "objects" at the temperature of the concerned medium, approximately 300 K. This gives us immediately     E ≈ $3.10^{-2}$ eV ≈ $5.10^{-21}$ J .

- The corresponding value of the action is therefore

$$A = L\sqrt{E \cdot M} = 10^{-10}\sqrt{4 \cdot 10^{-26} \cdot 5 \cdot 10^{-21}}$$

Or finally   $A \approx 14 \cdot 10^{-34}$ J.s $\approx 2h$ . This shows

- that it is strictly impossible – and in an unquestionable way illicit – to hope to formulate a realistic "classical" theory of membrane electrodynamics.
- that any attempt to "explain" based upon such a theory will be unfounded and will lead inevitably to a dead end.
- that the ultimate resort to "ad hoc" hypotheses or models will remove any admissible explanatory value and very real predictive capacity, and will lead to insurmountable conceptual obstacles…. These are precisely the difficulties that "classical" electrokinetics, and the models which wish to represent it, encounter unceasingly.

We therefore have considered here a totally different approach to the mechanisms in question which accuracy seems undeniable.

It is no longer an ad hoc hypothesis, but a reminder of the obligation of living organisms to respect a fundamental characteristic of the quantum level at which are located mechanisms which it implements. The following will show that a simple calculation allows at least a first approach to the mechanisms at cause, and that the orders of magnitude obtained are totally compatible with a "biological Tunnel effect".

The value of energy found above is $E = 5 \cdot 10^{-21}$ J. It concerns the average kinetic energy of the ions, resulting from the molecular agitation of the medium at a temperature of 300K.

The corresponding average speed $v_m$ will be obtained from

$$\tfrac{1}{2} M \cdot v_m^2 = E \quad \text{from where} \quad v_m = (2E/M)^{\tfrac{1}{2}}$$

The numeric value obtained is     $v_m \approx 0{,}5 \cdot 10^3$ m.s$^{-1}$

What then is the value of the membrane potential $V_0$, energy which would possess an unit charge placed at the entry of a pore, considered the "entry" of the potential barrier. It is particularly the minimal energy which a charge of the same sign must possess in order for the probability of crossing the barrier of potential to no longer be zero.

In the case considered here, the transmembrane DDP has a value of at least $\approx 7 \cdot 10^{-2}$ Volt, and this energy equals $V_0 = e \cdot V = 1{,}6 \cdot 10^{-19} \cdot 7 \cdot 10^{-2} = 1{,}1 \cdot 10^{-20}$ J $\approx 2 E_{\text{ion}}$, or more twice the average energy of one of the ions to be transferred.

The barrier is therefore completely impenetrable in "classical" terms.
One therefore finds $v_c \geq 0{,}7 \cdot 10^3$ m/s for "critical" transfer speed.
One can therefore say that "at least" $5 \cdot 10^{-21}$ J would be "missing from" the ion in order for the transfer probability to have a value other than zero.

Let us then consider the interaction of the quantum object with the potential barrier. It is a general fundamental quantum rule which will characterize this interaction.

In its context (thermal molecular agitation) this object is not in a proper state of its kinetic energy, and it does not present proper values. The result is that E will produce a wide spectrum of energy, in which we will encounter among others the values $E > V_0$ which will translate into the existence of non-zero probability amplitudes, showing the possibility of "crossing" of the barrier.

One will notice besides that the "fluctuation" required here thus finds its "explanation" and its value in the law of distribution of speeds of the present ions; here, the probability of an "instant" value of the speed greater than the average speed is never zero; it thus supplies to the concerned ions a non-zero probability of crossing the barrier on the condition however that the Heisenberg relationship, which always presents restrictions, be respected… One will find there the fundamental characteristics of the tunnel effect, in particular the Hartman's mechanism explaining the brevity and extreme efficiency of the process [13].

While we in fact know the "height" of the barrier, we do not yet know its exact configuration. The approximate calculation presented here will however show that the "thickness" of the barrier does not seem to significantly diminish the factor of transmission of ions across the critical zone, thus justifying the quantum approach via a typical Tunnel effect.

The <u>fundamental relationship</u> characterizing the quantum space is always:
$$\Delta E \cdot \Delta t \leq h$$
where $\Delta E$ represents a quantum fluctuation of energy occurring in the time interval $\Delta t$, or therefore $\Delta t \leq h / \Delta E$

Therefore $E + \Delta E > V_0$ and $\frac{1}{2} m v^2 = (E + \Delta E - V_0)$

We have: $v = [(2/m) \cdot (E + \Delta E - V_0)]^{1/2}$ "classical mean speed" of transfer

By definition $a = v \cdot \Delta t$ "width" of the barrier.

But in fact the relationship

$$v \cdot \Delta t \geq a \quad \text{or} \quad [(2/m) \cdot (E + \Delta E - V_0)]^{1/2} \cdot h / \Delta E \geq a$$

must be respected.

If one considers the first term of the inequality to be a function of ΔE, one establishes that it goes to a <u>maximum</u> of $\Delta E = 2(V_0 - E)$ which finally gives

$$[(2/m) \cdot (V_0 - E)]^{1/2} \cdot a \leq h$$

If one uses the typical value $2(V_0 - E)$ for $\Delta E$, one arrives, with the numeric values presented above, at a temporal fluctuation of $\approx 5 \cdot 10^{-14}$ s, the transfer of the quantum ion-object across the barrier must occur within a maximum delay of $5 \cdot 10^{-14}$ s, ie. at a minimum "mean speed" of about 100 miles/s in a "classical" interpretation. None of the biochemical processes presented in the current transfer models could ensure the sequences of reactions indispensable to this transfer, and only a quantum process can ensure it : it will be the Hartman's Effect. We know indeed that the transmissivity of the barrier (the crossing delay) very rapidly becomes independent of its thickness, the particle being then transferred in the shape of a "waves batch" which obeys the usual rules (attenuation, filtering of high frequencies). The weak transmissivity of the barrier is then compensated by the <u>extreme rapidity</u> of the transfer, ensuring its indispensible efficiency.

The barrier may then behave like a "thick" barrier and treated as such from the tunnel effect point of view.

Under these conditions, the transmission factor (Tr) of this barrier is obtained by

$$(Tr) = 4 E (V_0 - E) / V_0^2 \cdot \exp -2 [(m/2 \cdot (V_0 - E) \cdot a^2]^{1/2} \cdot a$$

An expression in which the value of $[ \ldots ]^{1/2}$ is $\geq h$.

With a barrier width estimated as the thickness of the membrane or 7 nm, and the values already used for $V_0$ and E, one finds a transmission factor <u>very close to 1</u>,

This confirms for us that the tunnel effect will be able to constitute the essential mechanism of the quantum object transfer (here ion $Na^+$ for example) "across" the membrane. The other form of tunnel effect, "wave", considered further below, will "cooperate" with the first to ensure the considered transfer.

It is important to also remember that the mean free path of the quantum objects of dimensions considered here, and placed in a liquid space, is in the amount of a few nanometers, that is precisely the amount of the thickness of the cell membrane to cross. This mean free path thus <u>would not result a priori in any extra restriction</u> on the work of the Tunnel effect.

One must finally consider of <u>equal importance</u> and influence the "undulate" aspect of the objects described here, an aspect which is strictly inseparable from their "corpuscular" characteristic, with an unquestionable exercise of Hartman's effect. In the case of the sodium $Na^+$ ion, the De Broglie relationship provides the wavelength associated with the quantum object "ion Sodium":
$$\lambda = h / m v$$
The "average speed" v still needs to be identified.

As shown above, the contribution of the kinetic energy of the ion at its full energy only depends ordinarily on the thermal energy at the temperature of the medium, or $\sim 300$ K.

The speed already found is in the amount of $5 \cdot 10^2$ m.s$^{-1}$: this is of course a minimum speed for the reasons presented above (Hartman), and it will correspond to a <u>maximum</u> De Broglie wavelength

This gives us the following for the De Broglie wavelength:

$$\lambda \approx 3 \cdot 10^{-11} \, m$$

Such a value could mean that the "wavy" quantum object, here « waves batch » can "see" the structure surrounding it, and be "guided" and "propelled" across this structure, here the specific ionic "channel", therefore ensuring its transfer (Hartman's mechanism). Such a mechanism seems thus identical to the guided propagation of an electromagnetic wave in a structure of nature and dimensions adapted, necessarily <u>higher</u> than the wavelength of the signal to propagate, which is <u>always</u> the case with the ionic "channel". It in fact must be understood that "wave" and "corpuscle", "macroscopic" concepts, cannot <u>separately</u> define the nature of the quantum object at cause here, any more than they can separately define its behavior. The theoretical equations and the numeric values obtained therefore correspond to one or another aspect of a quantum object which is determined by <u>the particular experiment chosen to permit measuring and observing</u>.

One can therefore only say that the "transfer" of the quantum object will unconditionally obey its "particle-wave" <u>double nature</u>, which will determine the circumstances of this transfer.

This no longer involves an ad hoc hypothesis but the statement of a rule which no experimental fact has contradicted: **the strict obligation of living organisms to respect the fundamental characteristics of the quantum level at which the mechanisms which it implements are located, to preserve its existence and ensure its functioning.**

To date, it seems that the field of cellular biology is one of the last to "evade" quantum physics, despite the absence of any justifiable reason, perhaps because of the extraordinary complexity of the problem….

Radioactivity $\alpha$, the glorious ancestor of the Tunnel effect, greatly assailed in its time, long unexplained and inexplicable, opens a new path to an approach and a comprehension of Nature which life sciences can no longer and should no longer avoid…